# On the mechanism of the flow of polymers

W. I. Kartsovnik

*Kultur-, Ingenieur-und Wissenschaftsgesellschaft e.V. (KIW-Gesellschaft),*

*Bautzner Str. 20 HH, 01099 Dresden,   Deutschland*

V. V.  Pelekh

*ZAO "TWINS Tec", Bolshaja Ochakovskaja Str.44, Of.9, 119361 Moscow, Russia*

Abstract

A non-Newtonian flow of a polymer melt is discussed. The description of the exponential decrease of the apparent viscosity by the well-known Eyring formula with an activation energy reduction proportional to the shear stress does not take into account specific features of the polymeric structure. We propose to modify the description of the macromolecular flow mechanism by including conformational changes of the polymeric chains. The elasticity of a strained polymeric chain, having an entropy origin, can be the reason of the reduction of the activation energy for the transition of a molecular-kinetic unit of the chain into a new equilibrium state in the flow direction during the thermal fluctuations. In that case, the activation energy of the transition should decrease by a value proportional to the reversible high-elastic component of the shear deformation caused by the flow of the polymer.

A non-linear dependence of viscosity on the shear rate well-known in studies of flow phenomena in condensed matter is observed in most of polymeric media, i.e. melted or diluted polymers. It is manifested in a decrease of the apparent viscosity $\eta$ defined as a ratio between the shear stress $\tau$ and the shear rate $\dot{\gamma}$:

$$\eta = \tau/\dot{\gamma} \qquad (1)$$

According to the theory of rate processes [1], an elementary act of the liquid flow driven by an external force consists in a thermally activated hopping of a molecule from one stable position to a neighboring free place (vacancy). This hopping requires an activation energy $E_0$. Fig.1 shows schematically such a process for a model system consisting of molecular-kinetic units represented, for simplicity, by spherical particles.

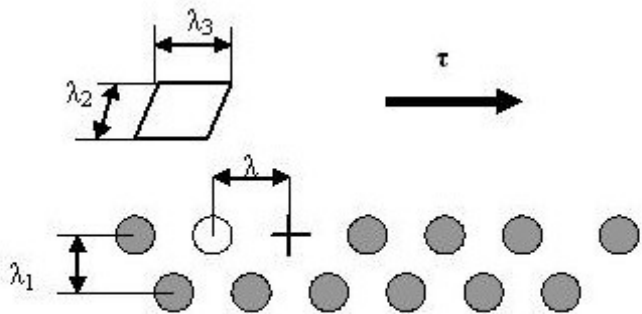

FIG. 1: A schematic illustration of a transition of a molecular-kinetic unit in a liquid from one stable position to another (H. Eyring [1]).

Considering a shear displacement of a molecular layer with respect to another layer, the viscosity can be defined as:

$$\eta = \tau\lambda_1 / \Delta u \qquad (2)$$

where $\lambda_1$ is the interlayer distance and $\Delta u$ is the velocity of one layer with respect to the adjacent one.

According to Eyring [1], a force applied to a liquid diminishes the height of the potential barrier for hopping in the direction of the force by the value $A = \frac{1}{2}\tau\lambda_2\lambda_3\lambda$. Here $\lambda_3$ and $\lambda_2$ - are the mean distances between two adjacent molecules of the moving layer in the direction parallel and perpendicular to the flow, respectively; $\lambda$ - is the distance between two equilibrium sites in the flow direction.

The hopping rate in this direction can be then expressed as:

$$k_f = \frac{k_B T F_a}{hF} e^{-(E_0 - A)/k_B T} \qquad (3)$$

where $k_B$ is the Boltzmann constant, $T$ is temperature, $F_a$ ($F$) is the sum of states of a molecule in the activated (initial) state per unit volume, $h$ is Planck's constant, and $E_0$ is the activation energy in the absence of external force.

In the opposite direction, the barrier height increases by the same value $A$ and the corresponding hopping rate is:

$$k_b = \frac{k_B T F_a}{hF} e^{-(E_0+A)/k_B T} \qquad (4)$$

Since a single hopping event shifts a molecule by the distance $\lambda$, the real flow velocity $\Delta u$ is:

$$\Delta u = \lambda(k_f - k_b) = \lambda k'(e^{-(E_0-A)/k_B T} - e^{-(E_0+A)/k_B T}) \qquad (5)$$

where $k' = \dfrac{k_B T F_a}{hF}$. Substituting Eq. (5) into Eq. (2), one obtains Eyring formula for the viscous flow:

$$\eta = \frac{\lambda_1 \tau}{\lambda k'(e^{-(E_0-A)/k_B T} - e^{-(E_0+A)/k_B T})} = \\ = \frac{\lambda_1 \tau}{2\lambda k' e^{-E_0/k_B T} \sinh(A/k_B T)}. \qquad (6)$$

If $A = \dfrac{1}{2}\tau\lambda_2\lambda_3\lambda \gg k_B T$, one can approximate:

$$\sinh\frac{A}{k_B T} \approx \frac{1}{2} e^{A/k_B T} \qquad (7)$$

and Eq. (6) can be rewritten as

$$\eta = \frac{\lambda_1 \tau}{\lambda k' e^{-(E_0-A)/k_B T}} \qquad (8)$$

Finally, assuming that $\lambda$ and $\lambda_1$ are of the same order, we rewrite Eq. (8) in the form:

$$\eta = B\tau e^{(E_0-\delta\tau)/k_B T} \qquad (9)$$

where $B = 1/k'$, and $\delta = \dfrac{1}{2}\lambda_2\lambda_3\lambda$ is the "viscosity volume" factor according to Eyring [1].
The dependence of viscosity on the shear stress in the form of Eq. (9) has been proposed by Eyring in the assumption that $\tau\lambda_2\lambda_3\lambda \gg 2k_B T$.

Formula (9) works fairly well in a relatively narrow range of deformation conditions. An extension of the shear rate range has revealed, however, considerable deviations from this dependence. To eliminate the problem, several attempts to generalize Eyring formula were made, by taking into account the presence, in anomalously viscous liquids, of molecular-kinetic units of different types characterized by different dimensions and relaxation time values. Many-parameter equations, such as Ree-Eyring can better describe the experimental data [2] but they all are empirical. The principal restriction of Eyring model is, however, that it considers a pure-viscous liquid, without taking into account such basic properties of polymers as visco-elasticity and high-elasticity. On the other hand, most of pure-viscous low-molecular liquids obey the Newton law of flow and display no anomalies of viscosity characteristic of practically all polymer melts and solutions.

Therefore, several models have been suggested in order to explain the viscosity anomalies from the viewpoint of the destruction of a thixotropical structured polymeric or any other anomalously-viscous system with increasing the shear rate and stress [3]. Specifically for polymer systems, of wide use have been concepts of net-like structures [4], hydrodynamic effects of rotating macromolecular spheres [5], non-linear visco-elasticity [6], allowing to fit experimental data by choosing appropriate operators and functions in rheological equations of state [7]. At present, there are many empirical formulae [8], which are able to describe one or another kind of experimental data.

We propose an alternative explanation of the exponential dependence of the viscosity of polymeric systems on the applied stress. To do that, we note that single molecular-kinetic units considered in Eyring model are, in fact, firmly bound with each other to units macromolecular chains. An external shear stress leads to stretching the chains [9] and therefore, as well known from the statistic theory of high-elasticity [10], to an increasing entropy elasticity.

Thus, one can expect that an atom of a macromolecule stretched along the flow direction experiences a force originating from the entropy elasticity of the chain. This force grows with increasing the deviation of the chain conformation from that corresponding to the thermal equilibrium. It is this force $\tau_e$, directly applied to the atom hopping into a new equilibrium position in the course of thermal fluctuations (see Fig. 2), which decreases the activation energy of the hopping.

Figure 2 displays in a schematic way the transfer of a molecular –kinetic unit of a polymeric chain (empty circle) into a vacancy (cross symbol). The expanded polymeric chain, tending to curl into a more stable, for the given temperature, conformation, is shown as black circles in Fig. 2. For simplicity, we assume, that the whole stretched macromolecule lies in layer 1, which moves under the external force with the rate $\Delta u$ with respect to the upper lying layer 2.

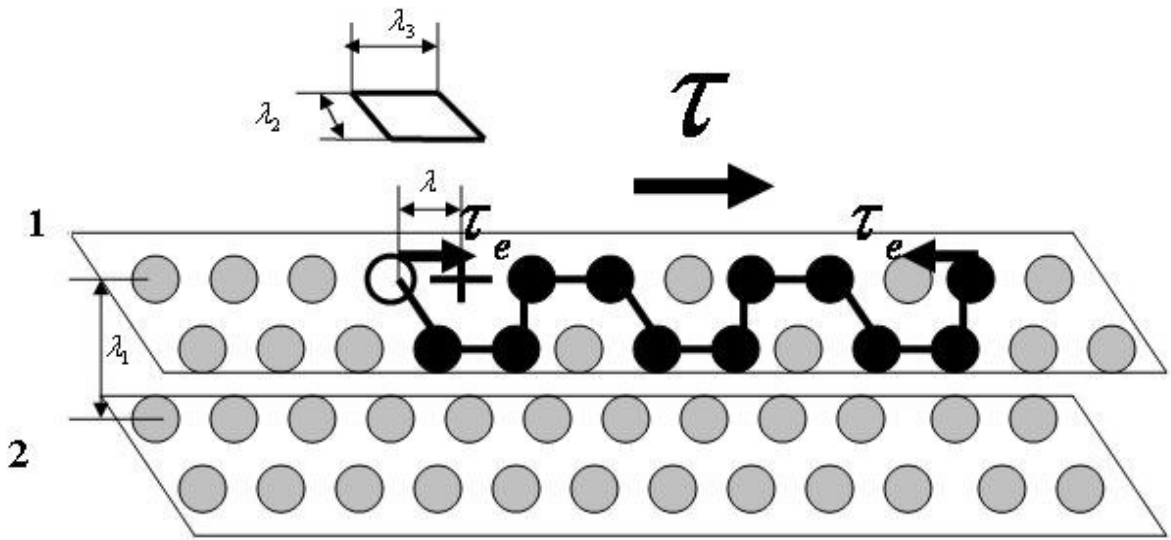

FIG. 2: A schematic illustration of a transition of a molecular-kinetic unit of a stretched polymeric chain into a new equilibrium position during the flow of two layers of the liquid with different rates with respect to each other

From the thermodynamic and statistic theory of high-elasticity it is known that the entropy elastic force *f* is proportional to the distance *r* between the terminal atoms of the macromolecular chain and to the temperature [11]:

$$f = \frac{3k_B T}{Nl^2} r \qquad (10)$$

where *N* is the number of the units in the chain and *l* is the length of a single chain.

An increase of the elasticity of caoutchouc at increasing temperature is an experimental confirmation of the entropy origin of the elasticity of macromolecules. However, investigations of deformation and flow processes in polymeric materials have shown that hitching and entanglement of macromolecular chains, i.e. fluctuating networks of physical bonds, are also important to take into account [6, 12, 13]. Such bonds and, especially, chemical bonds between macromolecules give rise to large reversible high-elastic deformations, also in flowing polymeric liquids and solutions [14, 15]. Conformational changes of large parts of the chain situated between the bonds, occurring during a chain deformation, lead to an increase of the entropy elasticity of the material. The statistic theory of high-elastic deformation of a three-dimensional network gives the following relationship between the shear stress $\tau_e$ and the shear deformation $\gamma_e$ of the network [11]:

$$\tau_e = nRT\gamma_e = G_e \gamma_e \qquad (11)$$

where $n$ is the number of the moles of the chains of the net in 1 cm$^3$ of the material, *R* is the universal gas constant, and $G_e$ is the shear modulus of the high-elastic network.

The flow rate of the upper molecule layer (layer 1), including units of the stretched polymeric chain (Fig. 2) with respect to the other molecule layer (layer 2) under the external shear stress $\tau$ is, according to Eq. (2), $\Delta u = \tau \lambda_1 / \eta$. This is the rate at which a molecular-kinetic unit of the stretched chain in the first molecule layer gets into the vacancy during the flow of the liquid. On the other hand, the hopping of atoms of the kinetic unit of the macromolecule to the nearest "hole" position is favored by entropy elasticity of macromolecule, stretched during the liquid flow. The reduction of the activation energy of the transition to a new equilibrium position (vacancy) occurs due to the force curling the stretched macromolecule, that is due to the shear stress of the high-elastic network, $\tau_e = G_e \gamma_e$. The sum of the rates of the transitions of the molecular-kinetic unit over the energy barrier into a new equilibrium position (into a vacancy) in the forward and backward directions is equal to the same flow rate $\Delta u$. Otherwise the macromolecule would start to curl or uncurl, repeatedly. At the steady-state flow with a given flow rate, the length of the macromolecule (or its part between the intermolecular bonds) in its uncurled form, which determines the high-elastic reversible deformation $\gamma_e$, is constant and provides a constant shear stress $\tau_e$ corresponding to the given flow rate $\dot{\gamma}$. Therefore, assuming that the activation energy is reduced by a value proportional to the entropy elasticity force $\tau_e$, and according to Eyring's analysis introduced above, we can write the following equation:

$$\eta = B\tau e^{(E_0 - \delta G_e \gamma_e)/k_B T} \qquad (12)$$

where $\tau$ is the shear stress measured during the flow, $G_e$ is the high-elastic shear modulus, and $\delta$ is a factor determined by the structure of the polymer and the shear conditions. It means that in the process of polymer's flow its viscosity $\eta$ depends exponentially on the value of its high-elastic reversible deformation $\gamma_e$.

This relationship can be verified using the experimental data on polyisobutylene flow curves measured by Vinogradov et al. [16]. In that work complete curves of polyisobutylene flow are given for five temperatures in a wide interval of the shear stresses $\tau$ and shear rates $\dot{\gamma}$. In addition, the reversible high-elastic deformations $\gamma_e$ of the polymer for the same shear stresses and rates have been measured after stopping the rotor of the viscosimeter. A strong anomaly in viscosity, consisting in its drop by a factor of 20, is also shown for the same data in Fig.5.2 of Ref. [8].

The applicability of formula (12) can be further checked by studying reversible high-elastic deformation in flowing polymer systems in a wide range of shear rates and stresses. For that, it is convenient to rewrite this expression [see Eq. (1) and (12)] as:

$$\frac{\eta}{\tau} = \frac{1}{\dot{\gamma}} = B \exp(E_0 - \delta G_e \gamma_e)/k_B T \tag{13}$$

Substituting $G_e = nRT$ (see Eq. (11)], and taking logarithm of both sides of Eq. (13), we arrive at:

$$\ln \frac{1}{\dot{\gamma}} = (\ln B + \frac{E_0}{k_B T}) - \delta' \gamma_e \tag{14}$$

where the factor $\delta' = \delta n \frac{R}{k_B}$ is determined by the slope of the dependence of $\ln(\frac{1}{\dot{\gamma}})$ on $\gamma_e$.

The work [16] presents the dependence of the high-elastic deformations of polyisobutylene on the logarithm of the shear rate at steady-state flow regimes for different temperatures. Recalculating this data to the dependence of the logarithm of the inverse shear rate on high-elastic deformation, we find a good linearity (see Fig. 5) in agreement with Eq. (14). The size $\delta$, determined by the slope of the dependence $\ln \frac{1}{\dot{\gamma}}$ on $\gamma_e$ in Fig. 3 by 22°C and $n = 10^{-4}$, is $4,4 \cdot 10^{-20}$ cm$^3$.

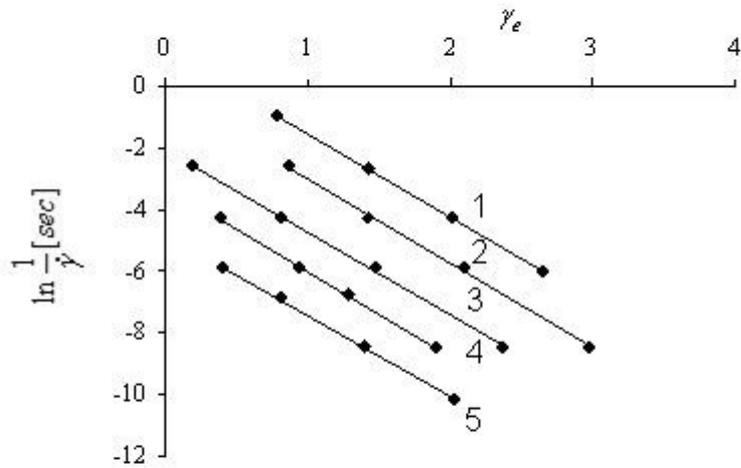

FIG. 3: Logarithm of the inverse shear rate vs. high-elastic deformations of polyisobutylene at steady-state flow regimes. Temperatures, °C: 1-22; 2-40; 3- 60; 4-80; 5-100. (Data recalculated from Ref. [16]).

The observed linear dependence of $\ln(\frac{1}{\dot{\gamma}})$ on high-elastic deformation $\gamma_e$ proves the validity of the proposed description of the anomalous viscosity of polymeric systems. Furthermore, the fact that the slope of this dependence keeps constant at different temperatures confirms the applicability of the entropy approach for describing the mechanism of the flow of polymers.

Finally, it should be noted that the present approach can be likely applied to flow processes in any condensed media whose structural elements cooperatively interact and influence atoms moving in the material.

*References*

[1] S.Glasston, K.J. Laidler, H. Eyring: *The Theory of Rate Processes* (New York and London, 1941).
[2] T. Ree, H. Eyring, J.Appl.Phys. **26,** №7, 793, 800 (1955).
[3] D.A.Denny, R.S. Brodkey, J. Appl.Phys. **33**, №7, 2269 (1962).

[4] W.W. Graessley, J.Chem.Phys. **43**, №8, 2696 (1965).
[5] F. J. Bueche, Chem.Phys. **2**, №9, 1570 (1954).
[6] P.J. Carreau, D.C.R. De Kee, R.P. Chabra, *Rheology of Polymer Systems, Principles and Applications* (Hanser/Gardner Publications, Inc. Cincinnati, 1997).
[7] T.W. Spriggs, Chem. Eng. Sci. **20**, №11, 931 (1965).
[8] G.V. Vinogradov, A.Ya. Malkin, *The Rheology of Polymer* (Khimiya, Moscow, 1977).
[9] M. Kroeger, Appl. Rheol. **5,** 66 (1995).
[10] L.R.G. Treloar, *The Physics of Rubber Elasticity* (Clarendon Press, Oxford, England, 1967).
[11] A.V. Tobolsky, *Properties and Structure of Polymers* (John Wiley & Sons, New York-London, 1960).
[12] H. Voigt, Appl. Rheol. **7**, 105 (1997).
[13] R.I. Tanner, A.M. Zdilar, S. Nasseri, Rheol. **44,** 513 (2005).
[14] G.V. Vinogradov, Mekh. Polim. № 1, 160 (1975).
[15] A.Ya. Malkin, Mekh. Polim. № 1, 173 (1975).
[16] G.V.Vinogradov, A.Ya. Malkin, V.F.Shumsky, Rheol. Acta **9,** 155 (1970).